\documentclass[12pt]{iopart}
\usepackage{epsfig}

\begin{document}

\title[Overview of Fluctuation and Correlation Results]{An Overview of Fluctuation and Correlation Results in Relativistic Heavy Ion Collisions}

\author{J T Mitchell\dag}
\address{\dag\ Physics Dept., Bldg. 510C, Brookhaven National Laboratory, Upton, NY 11973-5000 USA}

\begin{abstract}
A great deal of recent data on event-by-event fluctuation and correlation measurements has been released by several experiments at the SPS and RHIC. Recent results on charge fluctuations, balance functions in pseudorapidity, and average transverse momentum fluctuations will be reviewed. The results will be compared to various model predictions after examining contributions to each observable from known physics processes.
\end{abstract}


\section{Introduction}

Over the past several months, many results on event-by-event fluctuation and correlation observables have been made available by experiments at the SPS and at RHIC. This article will review the most recent results on three of these observables: net charge fluctuations, balance functions in pseudorapidity, and average transverse momentum fluctuations.  Each of these observables is designed to probe a specific property of the system: net charge fluctuations are sensitive to the charge distribution within the collision volume, balance functions are sensitive to the hadronization time of the system, and average $p_{T}$ fluctuations are sensitive to critical fluctuations of the temperature of the system which may be present during a phase transition from hadronic matter to a Quark-Gluon Plasma (QGP).  However, as with any measurement, the results can be contaminated by known processes such as charge conservation, resonance decays, elliptic flow, or even hard scattering processes.  Therefore, the contribution of any known processes must be thoroughly examined prior to formulating a reliable interpretation of the results. Detailed background information on the theory behind various event-by-event observables can be found in other recent reviews \cite{JeoRev}.

\section{Net Charge Fluctuations}

A promising signature for the presence of a QGP is the observation of event-by-event fluctuations in the net charge of produced particles.  It is hypothesized that fluctuations in the net charge would be significantly reduced in a QGP scenario.  This is due to the fact that in a QGP, the fractional electric charges of the quarks are more evenly spread throughout the QGP volume than the unit electric charges of the hadrons in a hadronic gas volume.  As a result, the fluctuations of the net charge in a given region of phase-space would be significantly reduced in a QGP \cite{Asa00,Jeo00}.

There has been much debate on the choice of the most appropriate observable to quote for the magnitude of net charge fluctuations. A standard is yet to be agreed upon in the literature, so PHENIX \cite{phnxq} quotes the normalized net charge variance, $v(Q)$, STAR \cite{starq} quotes a measure for dynamical fluctuations, $\nu_{+-,dyn}$, and NA49 \cite{na49qm} quotes the variable ${\phi}_{q}$. The latter will be quoted here.  Fortunately, all of the variables listed above can be related to each other with some very reasonable assumptions \cite{Pru02}.

The baseline for the measurement of net charge fluctuations is that of independent particle emission (where $v(Q)$ = 1.0, $\nu_{+-,dyn}$ = 0.0, and $\phi_{q}$ = 0.0). Predictions for the magnitudes of these three fluctuation variables for a variety of physical scenarios are listed in Table 1. Keep in mind that these values are valid only for small samplings of the total number of produced charged particles since charge conservation can contribute to any observed fluctuations. Net charge fluctuation measurements can be corrected for charge conservation based upon the fraction of the total number of particles in the sample as follows \cite{na49qm}:
\begin{equation}
{\phi}^{cc}_{q} = \sqrt{1 - (<N_{charged}>_{measured}/<N_{charged}>_{total})}.
\end{equation}

\begin{table}
\caption{\label{tab:1}The expected magnitude of various charge fluctuation
measures within specific physical scenarios for a small sample of the total
number of produced charged particles. $\nu_{+-,dyn}$ values are listed
within both the STAR and PHENIX acceptances.}
\begin{indented}
\item[]\begin{tabular}{@{}lllll}
\br
Scenario&v(Q)&$\nu_{+-,dyn}$ (STAR)&$\nu_{+-,dyn}$ (PHENIX)& $\phi_{q}$\\
\mr
Independent Emission&1.0&0.0&0.0&0.0\\
Resonance Gas&0.75&-0.0013&-0.006&-0.125\\
Quark Coalescence&0.83&-0.0008&-0.004&-0.084\\
Quark-Gluon Plasma&0.25&-0.0038&-0.019&-0.375\\
\br
\end{tabular}
\end{indented}
\end{table}

Figure 1 compiles the results of net charge fluctuation measurements by NA49 in $\sqrt{s_{NN}}$ = 17 GeV Pb+Pb collisions \cite{na49qm},  and by PHENIX \cite{phnxq} and STAR \cite{starq} in $\sqrt{s_{NN}}$ = 130 GeV Au+Au collisions, all quoted in terms of the $\phi_{q}$ variable.  Not included are new results from the CERES experiment \cite{ceresqm}. Superimposed on each plot are curves representing the predictions for independent particle emission, quark coalescence \cite{Bia02}, a resonance gas \cite{Jeo00}, and a QGP \cite{Jeo00}, after correcting for charge conservation within the acceptance of each detector.  At SPS energies, NA49 observes fluctuations that are most consistent with the independent particle emission scenario.  The PHENIX measurement is also consistent with independent particle emission.  However, for a large pseudorapidity window, STAR reports fluctuations that are most consistent with the prediction of the quark coalescence model.  Differences in the PHENIX and STAR measurements may be attributed to the reduced sensitivity of the PHENIX measurement, which may be improved in the analysis of Au+Au data at $\sqrt{s_{NN}}$ = 200 GeV.  Note that all of the measurements differ dramatically from the expectations for the pure QGP scenario, although it may be premature to rule out the possibility of the existence of a QGP based upon these results until more detailed models are consulted.

\begin{figure}[htb]
\vspace*{-0.12in}\begin{center}
\hspace*{-0.2in}
\includegraphics[scale=0.8,angle=0]{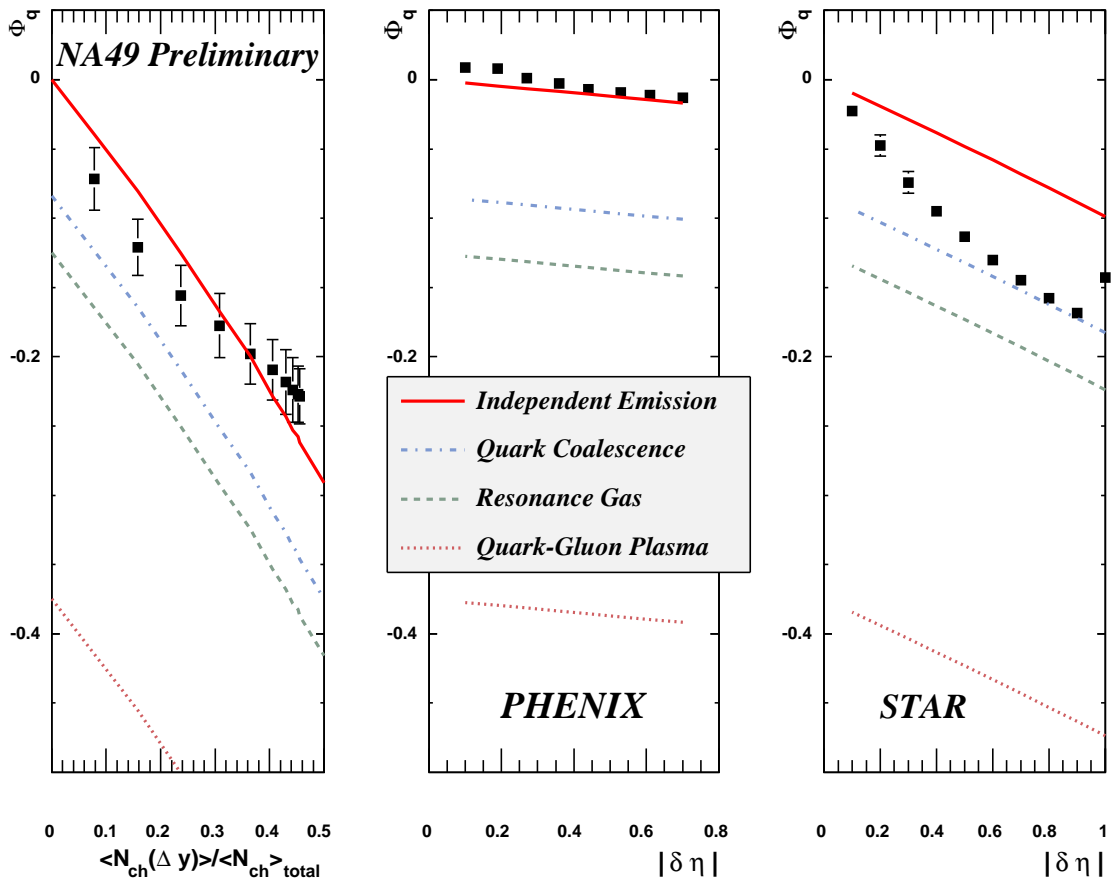}
\end{center}\vspace*{-0.25in}
\caption[]{Charge fluctuation measurements as a function of acceptance for a) $\sqrt{s_{NN}}$ = 17 GeV Pb+Pb collisions from NA49 \cite{na49qm}, b) $\sqrt{s_{NN}}$ = 130 GeV Au+Au collisions from PHENIX \cite{phnxq}, and c) $\sqrt{s_{NN}}$ = 130 GeV Au+Au collisions from STAR \cite{starq}. The curves represent expectations for independent particle emission, a resonance gas, quark coalescence, and a QGP after correction for charge conservation effects.  \label{fig:1}}
\end{figure}

\section{Balance Functions}

The balance function analysis method is designed to determine whether hadronization from a system occurs early ($<1$ fm/c) or at a later time \cite{Bas00}. The method is based upon the fact that charge is conserved locally due to produced pairs that are initially correlated in coordinate space. Concentrating on balance functions in pseudorapidity, the balance function is defined as: 
\begin{equation}
B(\Delta \eta) = \frac{1}{2} [\frac{N_{+-}(\Delta \eta) - N_{++}(\Delta \eta)}{N_{+}} + \frac{N_{-+}(\Delta \eta) - N_{--}(\Delta \eta)}{N_{-}}]
\end{equation}
Here, $N_{+-}({\Delta}\eta)$ refers to a histogram of $|\eta(h^{+}) - \eta(h^{-})|$ for all possible pairs of the subscripted charge signs within a single event, summed over all events.  $N_{+}$ and $N_{-}$ are the total number of positive and negative charged particles in the event, respectively. When examining the pseudorapidity of produced pairs if hadronization occurs at an early time, the pairs have more time to separate in $\eta$ due to a combination of the expansion of the system and rescattering.  Therefore, if there is a delayed hadronization scenario (as might be expected in central collisions), the analysis would yield a narrower balance function than the early hadronization scenario (as might be expected in more peripheral collisions).

Balance functions in pseudorapidity have been measured by STAR \cite{starbal,starbal04} and NA49 \cite{na49qm04}.  The behavior of the balance functions are studied as a function of centrality by plotting the widths of the balance functions in pseudorapidity ($<\Delta \eta>$) obtained from a Gaussian fit or from a weighted mean in ${\Delta} \eta$.  A reduction of the magnitude of $B({\Delta} \eta)$ for small ${\Delta} \eta$ is observed due to the HBT effect, but it has little effect on the extracted widths.  Figure 2 compares the balance function widths as a function of centrality at both SPS and RHIC energies. In both cases, there is a significant narrowing of the balance function observed for more central collisions.  STAR has shown that the widths in the most central collisions can be reproduced by a thermal model, but only if an unrealistically high value of the transverse flow velocity, $\beta_{T}=0.77c$ is used with a low temperature of 105 MeV \cite{starbal}. Taking the interpretation at face value, this behavior is indicative of a late hadronization scenario.

However, contributions from other known sources must first be considered when interpreting the results.  An obvious contributor to the width of balance functions are neutral resonance decays, which will register in the balance function as produced particle pairs that occur at a late time, serving to broaden the function.  The magnitude of resonance decay contributions using a thermal model baseline with $\beta_{T} \approx$ 0.5c has been investigated by P. Bozek et al. \cite{Boz03}. The results of the calculation can reproduce widths as a function of centrality that are consistent with the STAR data.  The behavior of balance functions as a function of centrality has also been investigated in the context of the quark coalescence model \cite{Bia03}.  This model can also reproduce the centrality-dependence and magnitude of the widths of the STAR data with $\beta_{T} \approx$ 0.5c, but without the inclusion of resonance decay contributions.

\begin{figure}[htb]
\vspace*{-0.12in}\begin{center}
\hspace*{-0.2in}
\includegraphics[scale=0.8,angle=0]{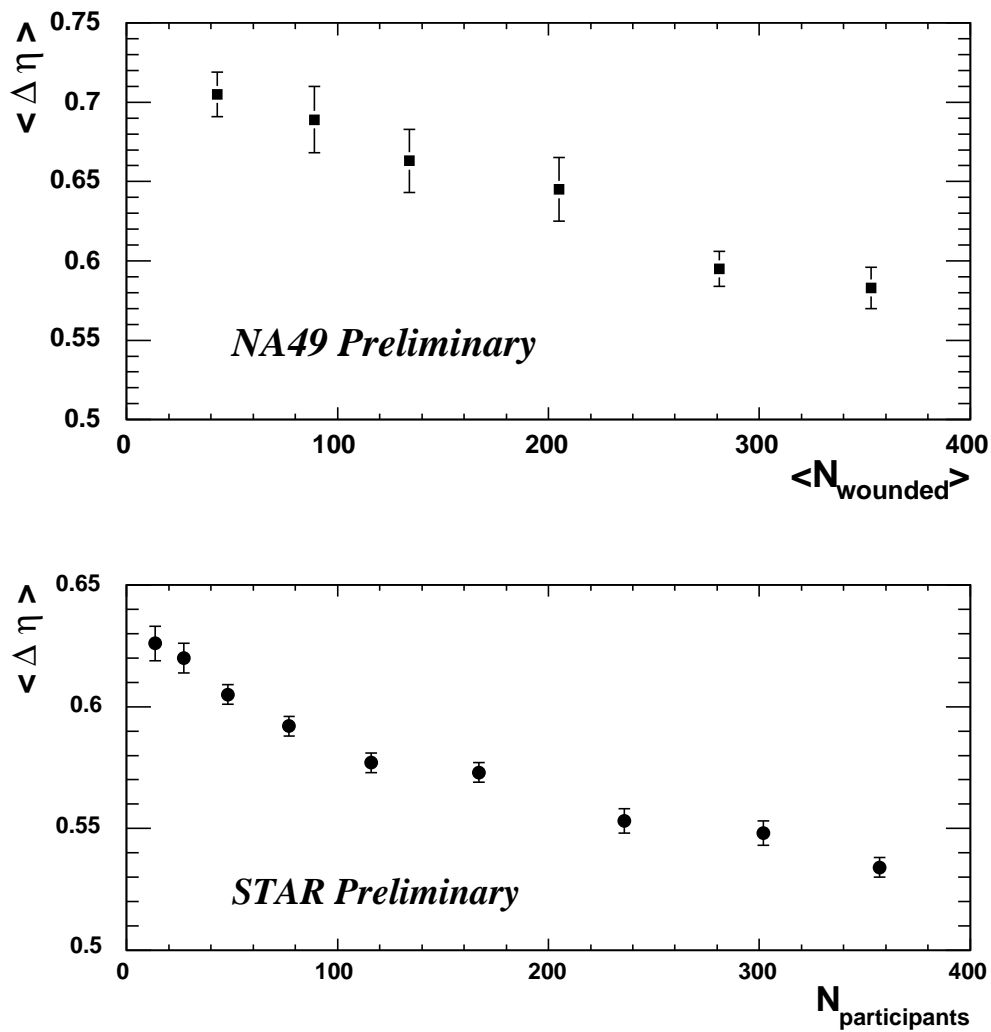}
\end{center}\vspace*{-0.25in}
\caption[]{The widths of balance functions in pseudorapidity ($<\Delta \eta>$) as a function of centrality for a) $\sqrt{s_{NN}}$ = 17 GeV Pb+Pb collisions from NA49 \cite{na49qm04}, and b) $\sqrt{s_{NN}}$ = 200 GeV Au+Au collisions from STAR \cite{starbal04}. In both cases, the balance functions are narrower for more central collisions. \label{fig:2}}
\end{figure}

\section{Average $p_{T}$ Fluctuations}

Event-by-event fluctuations in the average transverse momentum may be sensitive to temperature fluctuations near the QCD tri-critical point \cite{Ste99}.  There have been many recent results from the SPS (NA49 \cite{na49pt,na49qm04} and CERES \cite{cerespt,ceresqm}) and RHIC (PHENIX \cite{phnxpt,phnxqm04} and STAR \cite{starpt,starbal04}) experiments.  All experiments report a positive signal of non-random fluctuations, each with an apparent maximum in mid-peripheral collisions, as shown in Figure 3.  As with the charge fluctuation measurements, a standard observable has not been agreed upon in the literature, and each experiment quotes a different variable to describe the magnitude of the fluctuations: $\phi_{p_T}$, $F_{p_T}$, $\sigma^{2}_{dynamic}$, and $\Sigma_{p_T}$.  Fortunately, the variables can all be related with very reasonable assumptions \cite{Gav03}.  All of the variables measure the difference in the width of the event-by-event average $p_T$ distribution, $M_{p_T}$, from a random expectation, which can be expressed as a Gamma distribution that is directly calculable using the inclusive $p_T$ and number of particles per event, N, distributions \cite{Tan01}.

Since a positive signal is observed and confirmed, contributions to the signal from known sources should first be studied.  The various analyses have addressed this with respect to contributions from HBT, resonance decays, and elliptic flow. The contributions from all of these sources have been found to be very small or negligible, each confirmed by independent analyses.  Another possible source to the signal is that from hard scattering processes, which would be expected to produce correlations in $p_T$ with a stronger contribution at higher $p_T$.  In this context, both PHENIX and CERES observe that the majority of the contribution to the non-random fluctuations occur at high $p_T$.  PHENIX observes a large 243\% increase in the signal when the maximum of the $p_T$ range over which the average $p_T$ is calculated increases from 1.0 to 2.0 GeV/c.  This very large increase cannot be attributed to number fluctuations since the number of particles increases by less than 15\% over that range.  

In order to investigate possible contributions due to jet production, PHENIX has applied a two-component model consisting of 1) a simulation of soft processes by reproducing the inclusive $p_T$ and N distributions as a function of centrality, and 2) a simulation of hard scattering processes by embedding PYTHIA jet events at a given rate per produced particle in step 1 (while strictly conserving the N distribution so as not to introduce unnecessary number fluctuations).  The simulation is performed as a function of centrality within the detector acceptance in two modes: a) with the PYTHIA event embedding rate kept constant for all centralities, and b) with the PYTHIA event embedding rate scaled by the nuclear modification factor, $R_{AA}$, for each centrality bin. The result of the simulation superimposed with the PHENIX data is shown in the bottom plot of Figure 3. The simulation with the $R_{AA}$-scaled rate agrees with the centrality-dependent (and also the $p_T$-dependent \cite{phnxpt}) data trends remarkably well.  Within this simulation, the decrease of fluctuations in the most peripheral collisions is attributed to the fact that the signal begins to compete with the magnitude of the number fluctuations, while most of the decrease of fluctuations in the most central collisions can be attributed to the onset of jet suppression. When the simulation is performed at $\sqrt{s_{NN}}$ = 130 GeV within the STAR acceptance with the same PYTHIA event embedding rate that reproduces the PHENIX data, agreement with the STAR more central data is achieved within errors. The simulation predicts a 15-20\% increase in the fluctuation signal from STAR at $\sqrt{s_{NN}}$ = 200 GeV. Although the jet production cross section is much reduced at SPS energies, it is non-zero, and there may also be a residual jet contribution present. The identical simulation performed at $\sqrt{s_{NN}}$ = 17 GeV within the NA49 acceptance can also reproduce the trend of the NA49 data as a function of centrality.

A promising method in which to directly compare the fluctuation measurements from all four experiments is to estimate the magnitude of any residual event-by-event temperature fluctuations present using the prescription outlined in \cite{Kor01}.  The most central fluctuation values thus correspond to temperature fluctuations of 1.8\% in PHENIX, 1.7\% in STAR, 1.3\% in CERES, and 0.6\% in NA49.  The residual temperature fluctuations are small and do not significantly increase between SPS and RHIC energies.

Alternative interpretations of the observed average $p_T$ signal have been presented. One such interpretation is that the signal may be due to the onset of thermal equilibrium within the system \cite{Gav03}. This model is successful in reproducing both the centrality-dependence of the fluctuation signal and the inclusive $<p_T>$ distributions from STAR and PHENIX.

\begin{figure}[htb]
\vspace*{-0.12in}\begin{center}
\hspace*{-0.2in}
\includegraphics[scale=0.8,angle=0]{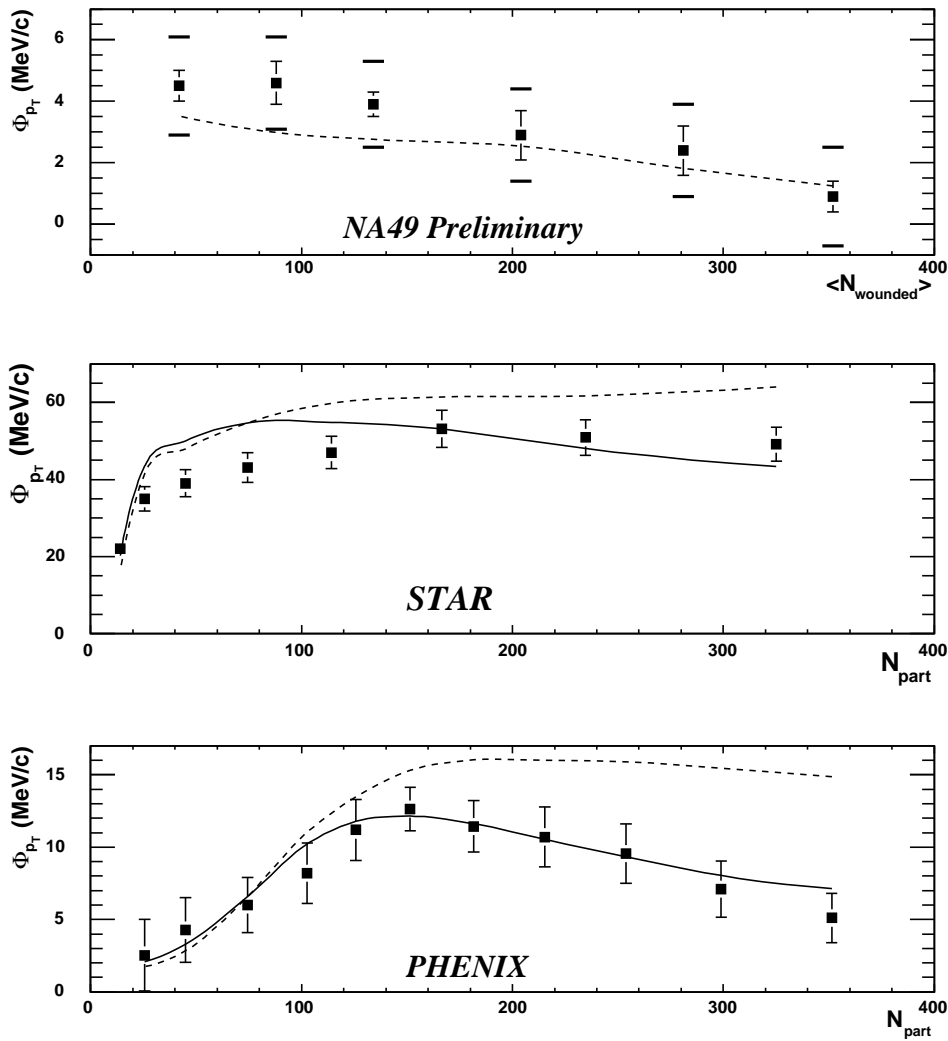}
\end{center}\vspace*{-0.25in}
\caption[]{Average $p_T$ fluctuations as a function of centrality for a) $\sqrt{s_{NN}}$ = 17 GeV Pb+Pb collisions from NA49 \cite{na49qm04}, b) $\sqrt{s_{NN}}$ = 130 GeV Au+Au collisions from STAR \cite{starpt}, and c) $\sqrt{s_{NN}}$ = 200 GeV Au+Au collisions from PHENIX \cite{phnxpt}. The error bars include statistical and systematic errors. The curves are the results of a model that estimates the relative contribution due to jet production as a function of centrality \cite{phnxpt} for a fixed PYTHIA event embedding rate (dashed lines) and for an $R_{AA}$-scaled embedding rate (solid lines).  Uncertainties on the simulation are 0.3 MeV/c for NA49, 2.1 MeV/c for STAR, and 1.0 MeV/c for PHENIX. \label{fig:3}}
\end{figure}

\section{Conclusion}

There has been nearly a doubling in the quantity of preliminary or published results on event-by-event fluctuation and correlation observables over the past year. Long thought of as the exclusive domain of large acceptance detectors, the scope of the results are demonstrating that event-by-event physics is accessible to smaller acceptance experiments with only a small sacrifice in sensitivity.  The development of event-by-event analysis techniques is continuing to mature at a rapid pace, including extensions to multi-dimensional correlations and methods to relate fluctuation observables directly to correlation functions.  It is becoming clear from what we have learned so far that these methods can provide additional tools with which to study the underlying dynamics of the collision system.

\end{document}